\def\be{\begin{equation}}
\def\ee{\end{equation}}
\def\bea{\begin{eqnarray}}
\def\eea{\end{eqnarray}}
\def\l{\label}
\def\c{\cite}

\def\massp{m_P}

\begin{document}
\begin{titlepage}

 \vspace{1.5in}

\begin{center}
\Large
{\bf A Potential Window to the GUT}

\vspace{.3in}

\normalsize

\large{James E. Lidsey}

\vspace{.7cm}

{\em NASA/Fermilab Astrophysics Center, \\
Fermi National Accelerator Laboratory, Batavia, IL 60510-0500. U. S. A. \\}


\end{center}

\vspace{5cm}

\baselineskip=24pt
\begin{abstract}
\noindent

The possibility that the form of the inflationary potential can be
reconstructed from a knowledge of the primordial power spectra of
scalar (density) and tensor (gravity wave) perturbations is discussed
and reviewed. It is suggested that measurements of the scalar spectral
index and the amplitude of the tensor spectrum at one scale should be
possible in the near future. This would be sufficient to determine
whether the potential is convex or concave over the limited region
relevant to the formation of large-scale structure in the universe.
Such information would provide an `observation' of physics at energy
scales above $10^{14}$GeV.

\vspace{2cm}

\noindent $^*$Invited talk presented at {\em The Second Alexander
Friedmann International Seminar on Gravitation and Cosmology, St.
Petersburg, Russia, 12th-19th September 1993.}

\vspace*{7pt}
\noindent
\small $^1$email: jim@fnas09.fnal.gov

\end{abstract}

\end{titlepage}

\section{Introduction}

\setcounter{equation}{0}

\def\theequation{\thesection.\arabic{equation}}

Einstein described the introduction of a cosmological constant into
the field equations of general relativity as `the biggest blunder of
his life.' In view of this, the idea that such a term might play an
influential role in the history of the universe has proved remarkably
popular. Theoretical cosmologists have traditionally looked to such a
term when attempting to reconcile theory with observation and indeed
Einstein's original motivation for considering a $\Lambda$-term was to
cancel the theoretically predicted expansion of the universe
\c{E1917}. Some time later the steady-state scenario was developed to
account for the apparently low age of the universe deduced from
observations of the expansion rate \c{HN1963}. In recent years the
inflationary scenario has been developed in an attempt to resolve some
of the puzzles of the hot big bang model \c{G1981}. Inflation explains
why the universe appears remarkably homogeneous and spatially flat on
large scales and why monopoles formed during phase transitions are not
observed at the present epoch.

During inflation the universe is dominated by the self-interaction
potential energy $V(\phi)$ of a quantum scalar field $\phi$ and this
false vacuum energy plays the role of an effective cosmological
constant for a finite time interval. This vacuum energy leads to a
negative pressure and hence gravitational repulsion. In this scenario
the matter content of the observed universe is formed when the false
vacuum decays at the end of inflation. Historically, Gliner \c{G1966}
first realized that a positive vacuum energy density is equivalent to
a cosmological constant and this idea was extended by Zel'dovich
\c{Z1968}. However, the idea that matter came from the expansion of a
universe dominated by a cosmological constant can be traced to a paper
by McVittie \c{M1952}. He found that a negative pressure in general
relativity can be converted into matter due to the expansion of the
universe if the quantity $\rho +3p$ is constant, where $\rho$ and $p$
are the energy density and pressure of the matter sector. This
quantity is approximately constant during inflation.

Current estimates suggest that over 1600 papers have been published on
various aspects of the inflationary scenario since its proposal by
Guth and others in 1981 \c{G1981}. However, despite such extraordinary
effort there remain some unresolved problems with the scenario. From a
particle physics viewpoint one of the most pressing problems with
inflation is the identity of the scalar field. Essentially the problem
is that there are too many candidates. These include the Higgs bosons
of grand unified theories, the extra degrees of freedom associated
with higher metric derivatives in extensions to general relativity, a
time-varying cosmological constant and the radius of the internal
space in Kaluza-Klein theories (for a recent review see \c{L1992}).

In general this scalar field is loosely referred to as the {\em
inflaton}.  Traditionally one chooses a specific model of particle
physics and then compares the theoretical predictions of the model
with observation. This enables the region of parameter space
consistent with observation to be determined. However, in view of the
large number of plausible models currently on the market, and
motivated by recent advances and results in observational cosmology,
one might ask whether observations of large-scale structure in the
universe can be employed to `reconstruct' the particle physics and in
particular the functional form of the inflationary potential [8-10].
We know from the observed quadrupole anisotropy of the cosmic
microwave background radiation (CMBR) \c{S1992} that the final stages
of the inflationary epoch occurred at or below the grand unified (GUT)
scale and this method therefore provides a potential window on the
physics of the early universe at scales of order $10^{16}$GeV. This
corresponds to $10^{-35}$s after the moment of creation.  The purpose
of this talk then is to review the reconstruction procedure and
discuss whether such a suggestion is feasable.

The different inflationary models may be classified into two main
groups.  Those in the first end via completion of a strong first-order
phase transition, whilst those in the second end when the scalar field
moves from a relatively flat region to a steeper region of its
potential (i.e. the slow-roll to waterfall transition).  In many cases
first-order inflation models can be expressed as slow-roll models
after an appropriate conformal transformation on the metric tensor and
in this talk we shall therefore adopt the working hypothesis that
inflation was driven by a single, self-interacting scalar field
minimally coupled to Einstein gravity. We further assume that the
initial conditions were specified by some appropriate quantum theory
of gravity at the Planck scale. In general this is the simplest
scenario and we shall refer to it as {\em basic inflation}. It is
certainly possible that a more complicated theory might be required to
accurately describe the physics of the very early universe, but at
this stage in the game we should at least attempt to rule out the
simplest models first before considering other possibilities further.

\section{The Formation of a Primordial Power Spectrum}

\setcounter{equation}{0}

In reconstruction we do not choose a particle physics model as an
initial condition. Consequently we may assume {\em nothing} about the
functional form of the potential {\em except that it leads to an epoch
of inflationary expansion}. To proceed, therefore, we must first
consider how the dynamics of the scalar field can be determined in
full generality.  Only the last 60 $e$-foldings or so of inflationary
expansion have significant observational consequences and in a very
real sense this represents the final stages of the inflationary
expansion. It is therefore reasonable to assume that any initial
anisotropies and inhomogeneities in the topology of the space-time
have been smoothed out by this stage. For the purposes of
reconstruction, therefore, one may assume the
Friedmann-Robertson-Walker (FRW) metric. It follows that the energy
and momentum equations for a universe dominated by the inflaton field
are
\be
\l{energy}
H^2=\frac{\kappa^2}{3}\left(
\frac{1}{2}\dot{\phi}^2+V(\phi) \right)-\frac{k}{a^2}
\ee
 and
\be
\l{momentum}
2\dot{H}=-\kappa^2\dot{\phi}^2 +\frac{k}{a^2},
\ee
where
$a(t)$ is the scale factor of the universe, $H\equiv \dot{a}/a$ is the
expansion rate, a dot denotes differentiation with respect to cosmic
time $t$ and $\kappa^2 \equiv 8\pi \massp^{-2}$, where $\massp$ is the
Planck mass. The spatial sections of the space-time are open, flat or
closed for $k=\{-1,0,+1\}$ respectively and we choose units such that
$\hbar =c=1$.

The field equation for the scalar field represents the local
conservation of energy and momentum. It follows directly from the
geometrical property that the boundary of a boundary is identically
zero:
\be
\l{bianchi}
\ddot{\phi}+3H\dot{\phi}+V'(\phi)=0,
\ee
where a
prime denotes differentiation with respect to $\phi$. The time
dependence in the energy equation may be eliminated by rewriting the
scalar field equation in terms of the energy density $\rho \equiv
\frac{1}{2} \dot{\phi}^2+V$ \c{L1991}. If $\dot{\phi}$ does not pass
through zero during the interval of interest (i.e. the field does not
oscillate), Eq. (\ref{bianchi}) simplifies to
\be
\l{bianchisimple}
\rho' =-3H\dot{\phi}, \qquad \dot{\phi} \ne 0.
\ee
The general
solution to this equation is
\be
t= -3\int^{\phi} d\phi' H(\phi')
\left( \frac{d\rho}{d\phi'} \right)^{-1}
\ee
and implies that the
inflaton field may be employed as an effective time coordinate. It
follows that $6H^2=-\rho ' X'/X$, where $X(\phi) \equiv a^2(\phi)$,
and the energy equation reads
\be \rho 'X'+2 \kappa^2 \rho X =6k.
\ee
The field equations reduce to the remarkably simple form
\be
\l{k=0}
2H'a'=-\kappa^2Ha, \qquad 2H' =-\kappa^2 \dot{\phi},
\ee
for $k=0$ and
substitution of Eq. (\ref{k=0}) into Eq. (\ref{energy}) determines the
potential via the Hamilton-Jacobi differential equation \c{SB1990}:
\be
\l{HJ}
(H')^2-\frac{3}{2}\kappa^2 H^2 =-\frac{1}{2} \kappa^4
V(\phi).
\ee

For reasons of technical simplicity, inflation is often discussed
within the context of the slow-roll approximation, but the
Hamilton-Jacobi formalism allows the dynamics of the inflaton field to
be investigated in full generality. Indeed, this framework has further
applications within the context of higher-order and scalar-tensor
gravity theories which are conformally equivalent to Einstein gravity
minimally coupled to a self-interacting scalar field \c{LidseyGRG}.

One may define the two parameters \c{LL1992}:
\be
\l{epsilon}
\epsilon
\equiv 3\frac{\dot{\phi}^2/2}{V+\dot{\phi}^2/2}=\frac{2}{\kappa^2}
\left( \frac{H'}{H}\right)^2
\ee
\be
\l{eta}
\eta \equiv
-\frac{\ddot{\phi}}{H\dot{\phi}} = \frac{2}{\kappa^2}\frac{H''}{H} =
\epsilon -\frac{\epsilon'}{\sqrt{2\kappa^2\epsilon}}.
\ee
Modulo
constants of proportionality $\epsilon$ is a measure of inflaton's
kinetic energy relative to its total energy density and $\eta$
measures the ratio of the field's acceleration relative to the
friction acting on it due to the expansion of the universe. We shall
refer to them as the {\em energy} and {\em friction} parameters
respectively.  The slow-roll approximation applies when the magnitudes
of these parameters are small in comparison to unity, i.e. $\{
\epsilon , |\eta | \} \ll 1$. Inflation proceeds when $\ddot{a}>0$ and
this is equivalent to the condition $\epsilon <1$. It is interesting
that only $H(\phi)$ and its first derivative determine whether the
strong energy condition is violated and in principle inflation can
proceed if $|\eta |$ is very large. This is the case if the field is
located within the vicinity of a local maximum in the potential.

This provides us with sufficient background to determine the
perturbation spectra.  Let us briefly review how primordial
fluctuations are generated during inflation \c{BST1983}. Density
perturbations, $\delta \rho$, arise after the field has rolled down
the potential well. Quantum fluctuations in the field during inflation
produce a time shift in how quickly the rollover occurs, thus
producing a $t\ne {\rm constant}$ hypersurface for $\delta\rho={\rm
constant}$. In other words, for a $t={\rm constant}$ 3-surface, there
is a density distribution produced by the kinetic energy of the
inflaton field.

In a universe with density field $\rho({\bf x})$ and mean density
$\rho_0$, the density contrast is defined as $\delta({\bf
x})=\delta\rho ({\bf x})/\rho_0 = (\rho ({\bf x})-\rho_0)/\rho_0$.
This contrast is most conveniently expressed as a Fourier expansion
$\delta ({\bf x}) \propto \int d^3k \delta_k \exp (-i {\bf k.x})$,
where we ignore the constant of proportionality. The density
perturbation on a scale $\lambda$ is then given by
\be
\left(
\frac{\delta\rho}{\rho} \right)^2_{\lambda} \propto \left. {k^3
|\delta_k|^2} \right|_{\lambda = k^{-1}}.
\ee

If the strong energy condition is violated, physical length scales
grow more rapidly than the Hubble radius $H^{-1}$, so a given scale
that starts sub-Hubble radius can pass outside the Hubble radius
during inflation and reenter after inflation during the radiation- or
matter-dominated epochs. This implies that quantum fluctuations
associated with a given length scale will be present when that length
scale reenters the Hubble radius. The amplitude of this fluctuation
when it reenters after inflation is given by
\be
\l{AS}
\left(
\frac{\delta\rho}{\rho} \right)^{\rm hor}_{\lambda} \equiv
\frac{m}{\sqrt{2}} A_S(\phi) =\frac{m\kappa^2}{8\pi^{3/2}}
\frac{H^2(\phi)}{|H' (\phi)|} ,
\ee
where the quantities on the
right-hand-side are to be evaluated when the scale $\lambda$ crossed
the Hubble radius during inflation \c{BST1983}. In the uniform Hubble
constant gauge the constant $m$ equals 4 or 0.4 if the fluctuation
reenters during the radiation- or matter-dominated era respectively.

In a similar fashion quantum fluctuations in the graviton field are
also redshifted beyond the Hubble radius during inflation. The
gravitational wave (tensor) spectrum is calculated in the
transverse-traceless gauge, where $+$ and $\times$ denote the two
independent polarization states of the metric perturbation. The
classical amplitude of the fluctuation satisfies the massless
Klein-Gordon equation, so the graviton behaves as a massless,
minimally coupled scalar field with two degrees of freedom
$\psi_{+,\times }$. The spectrum of tensor fluctuations is then given
by
\be
\l{AG}
A_G(\phi) = \frac{\kappa}{4\pi^{3/2}}H(\phi) ,
\ee
where
once again the quantities on the right-hand-side are evaluated when
the scale first crosses the Hubble radius during inflation \c{AW1984}.

The amplitude of the scalar fluctuations at horizon crossing is
related to the primordial power spectrum $P(k)$ via the relationship
$P(k) \propto A_S^2(k)k \propto k^{n(k)}$, where the function $n(k)$
defines the {\em spectral index}. The spectral indices of both the
scalar and tensor fluctuations may be expressed in terms of the energy
and friction parameters:
\be
\l{n}
1-n\equiv d\ln [A_S^2 (\lambda )]/d
\ln \lambda =2\left( \frac{ 2\epsilon_* -\eta_*}{\epsilon_*-1} \right)
\ee
\be
\l{nG}
n_G \equiv d\ln [A_G^2 (\lambda )]/d\ln \lambda =
\frac{2\epsilon_*}{1- \epsilon_*},
\ee
where $*$ indicates that the
energy and friction parameters should be evaluated when a particular
scale first crosses the Hubble radius [9,15,18].

This concludes our review of scalar field dynamics and the formation
of primordial power spectra in the very early universe.  We shall now
summarize the observational predictions generic to basic inflation and
will then discuss the formalism that enables a reconstruction of the
inflationary potential to be made, at least in principle!

\section{Observational Predictions of Basic  Inflation}
\setcounter{equation}{0}

Since the discovery of anisotropic structure on the CMBR there has
been renewed interest in the possibility that observational tests of
inflation might be possible within the near future. It is therefore
constructive to summarize the most generic predictions of basic
inflation.

\indent {\bf Prediction I}: Inflation puts the `bang' into the big
bang. The standard hot big bang scenario assumes as an initial
condition that the universe is expanding at some arbitrarily early
epoch and then traces the history of the universe after that epoch. In
the inflationary scenario, however, the gravitational repulsion of the
false vacuum leads naturally to a quasi-exponential expansion of the
scale factor and the observed Hubble expansion at the present epoch is
therefore a prediction of all inflationary models.

{\bf Prediction II}: The density parameter $\Omega \equiv \rho /\rho_c
=8\pi G/3H^2$ measures the ratio of the energy density of the universe
to the critical energy density at a given epoch. When the strong
energy condition of general relativity is satisfied the energy
equation (\ref{energy}) implies that $\Omega =1$ is an unstable
equilibrium point. However during inflation the strong energy
condition is violated and $\Omega $ approaches unity exponentially
fast. Although $\Omega$ has been evolving away from unity since
inflation ended, it follows immediately that the more inflation there
is, the closer the density parameter is to unity at the present epoch.
If the onset of inflation is determined by the initial conditions at
the Planck epoch, the scenario therefore predicts that the current
value of the density parameter measured on the horizon scale should be
\be
\Omega_0 =1\pm 10^{-5},
\ee
where the error of $10^{-5}$ arises
from the quantum effects that are responsible for the generation of
the primordial power spectrum and is estimated from the observed
quadrupole anisotropy in the CMBR. (Models where $\Omega_0$ differs
significantly from unity can be constructed, but they require very
special, and rather unnatural, initial conditions \c{omega}.
Consequently they do not satisfy our rather subjective definition of
basic inflation).

This prediction provides one of the main observational tests of the
inflationary scenario. When combined with the constraints from
nucleosynthesis, which imply that the baryonic matter can contribute
at most $10\%$ of the critical density, it leads to the prediction
that at least $90\%$ of the observable universe must consist of
non-baryonic dark matter. The discovery of dark matter would provide
strong support for inflation, although of course failure to detect
such particles would not disprove the scenario!

{\bf Prediction III}: For completeness we mention that there should
not be a significant abundance of monopoles left over from some early
phase transition, because their energy density is exponentially
redshifted by the inflationary expansion. The question of whether
cosmic strings can form after inflation is as yet unresolved, although
they can do so if the reheating temperature is sufficiently high.
However we shall not pursue these questions further here.

{\bf Predition IV}: It is clear that the energy parameter determines
the ratio of the amplitudes of the scalar and tensor modes. For scales
that reenter after decoupling this ratio is given by
\be
\l{ratio}
\frac{A_G}{A_S} =\frac{\sqrt{2}}{\kappa}\frac{|H'|}{H}
=\sqrt{\epsilon}.
\ee
We see immediately that the requirement that
inflation occurs, i.e. $\epsilon <1$, implies that
\be
\l{predictionI}
A_G < {A_S} .
\ee
 An observation violating this condition at any scale would
immediately rule out the general class of models we are considering.

{\bf Prediction V}: Substitution of Eq. (\ref{ratio}) into the
definition of the tensor spectral index leads to the relationship
\be
\l{predictionIII}
\frac{1}{2}n_G (\lambda) = \frac{A_G^2
(\lambda)}{A_S^2(\lambda) -A_G^2 (\lambda)}.
\ee
This expression is
valid for an arbitrary inflationary potential and illustrates a
fundamental connection between the forms of the scalar and tensor
fluctuation spectra. At each scale $\lambda$ it relates the amplitudes
of the two spectra to the tilt of the tensor spectrum. In principle
these three quantities can be determined observationally and
potentially this equation provides a powerful discriminator of the
inflationary hypothesis. We shall therefore refer to it as the {\em
consistency} equation \c{CKLL}.

We see from Eq. (\ref{predictionIII}) that $\epsilon <1$ also leads to
the prediction that
\be
\l{predictionII}
n_G \ge 0.
\ee
This implies
that the amplitude of the primordial gravitational wave spectrum must
always {\em increase} with increasing wavelength. This follows because
the amplitude of the tensor fluctuations when they reenter varies in
direct proportion to the expansion rate of the universe when that
scale first crossed the Hubble radius during inflation. However, if
$k=0$, Eq. (\ref{momentum}) implies that $H$ is always decreasing and,
since the first scales to go superhorizon are the last to reenter,
this implies $n_G>0$.

Some comments are in order at this stage. The expressions (\ref{AS})
and (\ref{AG}) for the scalar and tensor fluctuations are only
strictly valid to lowest-order in the energy and friction parameters.
The same is true for the consistency equation. In other words we have
assumed that the slow-roll condition is valid in deriving these
equations and this goes against the spirit of reconstruction. This
assumption is necessary because the full analytical expressions for
the perturbation spectra are unknown at present. Improved expressions
valid to first-order in $\epsilon$ and $\eta$ are
\bea
\label{scalar}
A_S & \simeq & - \frac{\sqrt{2} \kappa^2}{8\pi^{3/2}} \,
\frac{H^2}{H'} \, \left[ 1 - (2C+1)\epsilon + C \eta \right] \\
\label{wave}
A_G & \simeq & \frac{\kappa}{4\pi^{3/2}} \, H \, \left[ 1
- (C+1) \epsilon \right] \, ,
\eea
where $C = -2 + \ln 2 + \gamma
\simeq -0.73$ is a numerical constant and $\gamma \approx 0.577$ is
the Euler constant \c{SL1993}.  However, for reasons of technical
simplicity, we shall employ the lowest-order expressions in the
remainder of this talk.

We now proceed to develop the formalism that allows the functional
form of the inflationary potential to be reconstructed from a
knowledge of the primordial power spectra of scalar and tensor
fluctuations.

\section{Recontruction of the Inflationary Potential from the
Primordial Power Spectra}

At this stage it is worth remarking that only a very small region of
the inflationary potential is available for reconstruction. Scales of
cosmological interest span the narrow range $1h^{-1}$Mpc to
$6000h^{-1}$Mpc, where $H_0 = 100h$ km ${\rm s}^{-1}$ ${\rm Mpc}^{-1}$
is the current expansion rate of the universe. These scales correspond
to galaxies through to the size of the current horizon. Such a range
of scales first crossed the Hubble radius during $\ln 6000 \approx 9$
$e$-foldings of accelerated expansion and this represents a very small
part of the total inflationary era.

We shall assume that the spectra $A_S(\lambda)$ and $A_G(\lambda)$
have been determined observationally, at least over a small range of
scales, by combining large-scale structure and CMBR experiments. The
reconstruction of the inflationary potential now follows by
parametrizing the full set of solutions in terms of the functional
$H[\lambda(\phi)]$. The expressions for the amplitudes of the scalar
and tensor fluctuations become
\bea
\label{spectra}
A_S(\lambda) & = &
\frac{\sqrt{2}\kappa^2}{8\pi^{3/2}} \, H^2(\lambda) \left|
\frac{d{\lambda}}{dH}\, \frac{d\phi}{d\lambda} \right| \nonumber \\
A_G(\lambda) & = & \frac{\kappa}{4\pi^{3/2}} \, H(\lambda),
\eea
respectively.

There exists a one-to-one correspondence between a given length scale
$\lambda$ and the value of $\phi$ when that scale crossed the Hubble
radius during inflation. The physical size of a scale at the present
epoch is simply $\lambda (\phi) =H^{-1}(\phi) a_0/a(\phi)$, where
$H(\phi)$ and $a(\phi)$ are determined at the epoch when the scale had
physical size $H^{-1}(\phi)$ (i.e. when it crossed the Hubble radius
during inflation). The value of $a(\phi)$ is related to the size of
the scale factor at the end of inflation, $a_e$, via the expression
\be
a(\phi)=a_e \exp [-N(\phi) ],
\ee
where
\be
N(\phi) \equiv
\int^t_{t_e} H(t')dt' =-\frac{\kappa^2}{2} \int^{\phi_e}_{\phi} d\phi
' H(\phi ') \left( \frac{dH(\phi ')}{d\phi '} \right)^{-1}
\ee
is the
number of $e$-foldings of growth from a particular value of $\phi$ to
the end of inflation at $\phi_e$ (defined in general as the point
where $\epsilon$ reaches unity).  This implies that
\be
\label{lambdaphi}
\lambda(\phi)=\frac{\exp[N(\phi)]}{H(\phi)} \,
\frac{a_0}{a_e},
\ee
and differentiation of this equation with respect
to the inflaton field yields
\be
\label{dlambdaphi}
\frac{d\lambda(\phi)}{d\phi}= \pm \frac{\kappa}{\sqrt{2}} \left(
\frac{A_S}{A_G} - \frac{A_G}{A_S} \right) \lambda .
\ee

An expression for the potential, as parametrized by the scale
$\lambda$, follows by substituting Eq. (\ref{spectra}) directly into
the Hamilton-Jacobi equation (\ref{HJ}):
\be
\label{Vlambda}
V[\phi(\lambda)] = \frac{16\pi^3A_G^2(\lambda)}{\kappa^4}\left[ 3 -
\frac{A_G^2(\lambda)}{A_S^2(\lambda)} \right] .
\ee
Finally, we need
to determine how $\lambda$ varies with $\phi$. Integration of Eq.
(\ref{dlambdaphi}) yields the function $\phi=\phi(\lambda)$ given by
\be
\label{philambda}
\phi(\lambda) = \pm \frac{\sqrt{2}}{\kappa }
\int^\lambda \frac{d\lambda'}{\lambda'}
\frac{A_S(\lambda')A_G(\lambda')}{A_S^2(\lambda')-A_G^2(\lambda')}.
\ee
An alternative form for $\phi (\lambda)$ may be derived by
substituting the consistency equation (\ref{predictionIII}) into Eq.
(\ref{philambda}):
\be
\label{phispectra}
\phi=\pm
\frac{\sqrt{2}}{\kappa} \int^{A_G} dA_G'\, \frac{A_S[A_G']}{A_G'^2 }.
\ee
Without loss of generality the arbitrary integration constant has
been eliminated by performing a linear translation on the value of
$\phi$. This second expression proves particularly useful in the
reconstruction process if the functional form of $A_S$ as a function
of $A_G$ is known.

The functional form $V(\phi)$ of the potential is deduced by inverting
Eq. (\ref{philambda}) and substituting into Eq. (\ref{Vlambda}). The
reconstruction equations are Eq. (\ref{Vlambda}) and (\ref{philambda})
along with the consistency equation (\ref{predictionIII}). We conclude
that {\em full reconstruction requires a knowledge of both the scalar
and tensor perturbation spectra}. However, recall that the tensor
spectral index is just $n_G=2d\ln A_G /d\ln \lambda$. If the tensor
spectrum is known, this implies that one may employ the consistency
equation to derive the scalars. Unfortunately the consistency equation
is a first-order, ordinary differential equation, so the reverse is
not true! If one only has a knowledge of the scalar spectrum (and from
an observational point of view this is the most likely scenario), one
must integrate the consistency equation to determine the tensors and
this necessarily introduces an arbitrary constant into the tensor
spectrum. The non-linear nature of the consistency equation implies
that the scalar spectrum alone does not uniquely determine the tensor
spectrum, and hence the functional form of the inflationary potential.

As a corollary, an arbitrarily accurate determination of the scalar
spectrum is insufficient to reconstruct the potential. {\em A minimal
knowledge of the primordial gravitational wave spectrum is required}.
Indeed the amplitude of the tensor spectrum at one scale is sufficient
to determine the integration constant and lift the degeneracy. We
shall discuss in Section 6 some of the possible observational routes
whereby the necessary information may be gathered.

\section{Reconstruction from a Constant Scalar Spectral Index}

\setcounter{equation}{0}

 In this section we shall derive the general class of inflationary
potentials that leads to a spectrum of scalar fluctuations with a {\em
constant} spectral index. We assume a power law of the form
\be
\l{powerlaw}
A_S(\lambda ) \propto \lambda^{(1-n)/2} \propto
k^{(n-1)/2}.
\ee
This class of spectra is the simplest extension to
the scale-invariant Harrison-Zel'dovich spectrum ($n=1$) and at any
rate it is likely that more complicated features in the spectrum will
not be measureable observationally in the near
future.\footnote{Throughout this talk the term `near future' refers to
any timescale contained within the next decade.} However, recent
results already constrain power spectra of the form (\ref{powerlaw}).
The data from COBE suggests that (\ref{powerlaw}) is consistent at the
1-$\sigma$ level if $n$ lies in the range $0.6$ to $1.6$. These limits
are independent of the dark matter components. If one includes
clustering data and assumes a cold dark matter (CDM) model, the lower
limit becomes $n > 0.7$ at $95\%$ confidence if gravitational waves do
not contribute significantly to the CMBR temperature anisotropy and $n
> 0.84$ if they do \c{LL1992}.

The functional forms of $H(\phi)$ that lead to such spectra may be
derived by equating Eqs. (\ref{AS}) and (\ref{powerlaw}), taking
logarithms, and differentiating with respect to the scalar field to
remove all undetermined constants \c{CL1993}. The $(\ln \lambda )'$
term may be rewritten after substitution of Eq. (\ref{dlambdaphi}) and
this leads to a second-order differential equation in $H(\phi)$
\c{CL1993}:
\be
\l{H''}
 2(5-n) \frac{(H^\prime)^2}{H} - 2H^{\prime\prime}=-(n-1) \kappa^2 H.
\ee
The order of this equation may be reduced by using the identity
\be
2H''\equiv \frac{d(H')^2}{dH}
\ee
and we arrive at the non-linear,
first-order differential equation
\be
\l{H'^2}
\frac{ d
(H^\prime)^2}{d H}- (5-n) \frac{H^{\prime^2}}{H} = \frac{n-1}{2}
\kappa^2 H.
\ee
This admits the exact integral
\be
\l{H'}
(H^\prime)^2= \frac{n-1}{n-3} \frac{\kappa^2}{2} H^2 +CH^{5-n}, \qquad
n \ne 3,
\ee where $C$ is an arbitrary integration constant.  Eq.
(\ref{H'}) has a number of solutions depending on the signs of $C$ and
$n$. We have summarized them in Table 1 and the second integration
constant has been absorbed into the value of the scalar field
\c{LT1993}.

\begin{table}
\begin{center}
\begin{tabular}{c||c|c|c}
$C/ n$& $n<1$& $n=1$& $n>1$ \\
\hline
\hline
         &          &          &            \\
$C<0$& $\Lambda {\rm sech}^m (\omega\phi)$& NS& NS\\
         &          &          &            \\
$C=0$& $\exp \left( \pm \sqrt{\frac{1}{2}\frac{n-1}{n-3}}\kappa\phi \right)$&
Const.& NS\\
         &          &          &             \\
$C>0$& $ \Lambda {\rm cosech}^m (\omega\phi)$&$\frac{1}{\sqrt{C}\phi}$
&$\Lambda
{\rm sec}^m (\omega\phi)$\\

\end{tabular} \end{center} \footnotesize{\hspace*{0.2in} Table 1: The
functional forms for $H(\phi)$ that lead to a primordial power
spectrum of scalar fluctuations with constant spectral index are
tabulated for positive, vanishing or negative $C$ and $n-1$.  The
parameter $\Lambda$ is a positive\---definite constant determined by
$C$ and `NS' implies no real solution for $H(\phi)$ exists in the
given region. $m=2/(3-n)$ and $\omega^2 = |(n-1)(n-3)|\kappa^2/8$ if
$n<3$.  The form of the potential follows from the Hamilton-Jacobi
equation (\ref{HJ}). We see that the integration constant $C$ must be
specified for the potential to be uniquely determined. This can only
be done at an observational level if one has knowledge of the
primordial gravitational wave spectrum on at least one scale of
interest.}

\vspace{.7cm}
\end{table}

The inflationary potential follows by substituting these solutions
into the Hamilton-Jacobi equation (\ref{HJ}). Its form can not be
determined uniquely unless the sign of the integration constant is
specified. This constant determines the energy scale at which the
fluctuations are first formed during inflation and therefore such a
specification requires knowledge of the gravitational wave spectrum on
at least one scale. To investigate this further let us consider the
$n<1$ solutions in more detail when $C$ is negative semi-definite.

Both the hyperbolic and exponential potentials lead to a constant
spectral index $n$. What distinguishes the two solutions is the
relative amplitude of the gravitational wave spectrum. This is
determined by the ratio (\ref{ratio}). The energy parameter in the two
cases is given by
\be
\l{C=0}
\epsilon = \left( \frac{n-1}{n-3}
\right)
\ee
and
\be
\l{C<0}
\epsilon = \left( \frac{n-1}{n-3} \right)
\left[ {\rm tanh} \left( \sqrt{ \frac{(n-1)(n-3)}{8}} \kappa \phi
\right) \right]^2
\ee
for $C=0$ and $C<0$ respectively. In the latter
case it can be shown that cosmological scales crossed the Hubble
radius during inflation when $|\kappa\phi|$ was very small, thereby
implying that the contribution of the gravitational waves is
exponentially suppressed in this model \c{L1993}.

It is worth remarking that the solutions presented in Table 1 are
general and therefore valid for all values of the scalar field. This
implies that one can always expand the potentials as a Taylor series
about some specific value $\phi_0$ and then perform a linear
translation on the field \c{Turner}. It is interesting to note that
the expansions for the $\{n<1,C<0\}$ and $\{n>1,C>0\}$ are both given
by
\be
\l{expansions} H(\phi) =\Lambda \left[1 +\left( \frac{n-1}{8}
\right) \kappa^2 \phi^2 + {\cal{O}} (\phi^4) \right] .
\ee
This
implies that an almost constant spectral index will always be obtained
if $H(\phi)$ has an expansion of the form (\ref{expansions}) to lowest
order in $\phi$. Such an expansion is valid because all large-scale
structure observations correspond to a very narrow region of the
inflationary potential and since the field must be rolling down its
potential at a sufficiently slow rate for inflation to proceed in the
first place, the relative change in the position of the field over the
$9$ $e$-foldings associated with large-scale structure observations is
expected to be tiny.

Potentials of this form arise in a number of particle physics models.
For $n<1$ the potential can be interpreted physically as a bulk
viscous stress acting on the energy-momentum tensor of a perfect
barotropic fluid \c{L1993}. The self-interaction potentials of the
pseudo-Nambu-Goldstone bosons also have the form (\ref{expansions}) in
the small angle approximation \c{natural...}. Furthermore, the
recently proposed hybrid inflationary scenario is driven by potentials
with an expansion given by (\ref{expansions}) and predicts $n>1$
\c{Linde}.

Finally we summarize the relationship between the tilt of the scalar
power spectrum and the gravitational wave contribution in Table 2
\c{BL1993}. The important quantities are the energy and friction
parameters. These are described as `large' when they are significantly
larger than zero but still less than unity and `small' when they are
very close to zero.

\begin{table}
\begin{center}
\begin{tabular}{c||c|c}
  Scalar      & Gravitational Waves        & Gravitational Waves     \\
  Spectrum    & Important                  & Negligible              \\
\hline
\hline
               &                           &                         \\
Small          & $\epsilon$ large                 & $\epsilon$ small
   \\
Tilt           & $2\epsilon \approx \eta$        & $|\eta|$ small          \\
               &                           &                         \\
Significant    & $\epsilon$ large                & $\epsilon$ small
 \\
Tilt           & $|\eta|$ large            & $|\eta|$ large          \\

\end{tabular} \end{center} \footnotesize {\hspace*{.3in} Table 2 -
Illustrating the connection between tilt and the gravitational wave
production in terms of the magnitudes of the energy and friction
parameters.} \end{table}

\section{Can the  Primordial Gravitational Waves be Observed?}
\setcounter{equation}{0}

In this section we shall assess the current possibilities for
reconstructing a part of the inflationary potential. It is clear that
at some level one requires knowledge of the gravitational wave
spectrum and there are three possible methods by which such a spectrum
may be determined.

{\em 1. Direct Detection}: The most obvious method is via direct
detection. For inflation driven by exponential potentials the spectral
indices satisfy $n_G=1-n$ and are uniquely determined by the energy
parameter via Eq. (\ref{C=0}). Hence, increasing the tilt increases
the contribution of the gravitational waves to the CMBR anisotropy on
large angular scales. However, tilting the spectrum reduces the
amplitude on the smaller scales relevant for direct detection.  The
predicted amplitude of the waves on scales relevant to the most
sensitive direct probes of gravity waves, such as the Laser
Interferometer Gravity-wave Observatories (LIGO), are too low to be
detected \c{gravy}. Consequently this route does not appear promising.

{\em 2. Polarization of the CMBR}: Polarization of the CMBR can arise
if the microwave radiation is scattered by free electrons due to
Thomson scattering in the presence of a gravitational wave. In
principle measurements of the CMBR polarization can provide valuable
information regarding the contribution of gravitational waves to
large-scale temperature anisotropies \c{CDS1993}. Recent numerical
calculations suggest, however, that such an effect is very difficult
to detect in practice, since the combined polarization of the tensor
plus scalar is typically less than $1\%$. Therefore this approach does
not seem particularly promising either.

{\em 3. Comparison of small and large scale CMBR anisotropies}: The
most promising method for determining the gravitational waves at a
particular scale is by comparing the large and small angle
anisotropies (for a discussion of these issues see \c{C}). The surface
of last scattering is located at a redshift $z_{\rm LSS} \approx 1100$
and the angle subtended by the horizon scale at that redshift is
approximately $\theta_{\rm LSS} \approx (1+z_{\rm LSS})^{-1/2} \approx
2^o$. This implies that anisotropy measurements on angular scales
above $\theta \approx 2^o$ directly determine the primeval form of the
fluctuations. On smaller angular scales the fluctuations have
reentered the horizon. After reentry the gravitational waves behave as
relativistic matter and their energy density redshifts as the fourth
power of the scale factor. This implies that only scalar modes affect
the CMBR anisotropy below $\approx 2^o$. If the fluctuations are
statistically independent and Gaussian-distributed, the scalar and
tensor fluctuations on scales above $2^o$ add in quadrature.
Consequently {\em a comparison of large and small scale CMBR
experiments may yield information regarding the influence of the
primordial gravitational waves}.

Let us investigate this further. The CMBR anisotropies can be expanded
into spherical harmonics \c{SV1990}
\be
\frac{\Delta T}{T} ({\bf
x},\theta,\phi) = \sum_{l=2}^{\infty} \sum_{m=-l}^{l} a_{lm}({\bf x})
\; Y^l_m(\theta,\phi) ,
\ee
where ${\bf x}$ denotes the observer's
position and the coefficients $a_{lm}({\bf x})$ are Gaussian random
variables with mean and variance determined uniquely by the harmonic
$l$:
\be
\langle a_{lm}({\bf x}) \rangle = 0 \; \; \; ; \; \; \;
\langle \left| a_{lm}({\bf x})\right|^2\rangle \equiv \Sigma_l^2 .
\ee
A given inflationary model predicts values for the averaged
quantities
\be
\langle Q_l^2 \rangle = \frac{1}{4\pi} (2l+1)
\Sigma_l^2 ,
\ee
where the average is taken over {\em all} observer
positions, but the observed multipoles measured from a single point in
space are given by
\be
 Q_l^2 = \frac{1}{4\pi} \sum_{m=-l}^{l} \left|
a_{lm} \right|^2 .
\ee

For sufficiently small angles on the CMBR sky the $l$th harmonic of
the expansion is given approximately by $l \approx (200^o/\theta)$.
Hence for harmonics smaller than $l \approx 20$ the dominant
contribution to the CMBR arises solely from the Sachs-Wolfe effect
when photons are either red or blue shifted as they climb out of, or
fall into, gravitational potential wells \c{SW1967}.  If one assumes a
constant spectral index $n<3$, the variances of the coefficients in
the harmonic expansion of the scalar fluctuations are related by
\be
\l{gammas}
\Sigma_l [S]=\Sigma_2 [S] \frac{\Gamma[l+(n-1)/2]}{\Gamma
[l+(5-n)/2]} \frac{\Gamma [(9-n)/2]}{\Gamma [(3+n)/2]}.
\ee
Hence, in
models where gravitational waves are not significant, measurements of
the lower-order harmonics provide a fit to the spectral index.

For models where the tensor fluctuations are important, however, the
anisotropies from such modes are
\be
\Sigma^2_l [T] = 144\pi^5 G(2l+1)
\frac{(l+2)!}{(l-2)!} \left( \frac{H_0}{2}\right)^{n-1} C(n)
\int^{\infty}_0 dk k^{n-2} I_l^2 (k),
\ee
where
\be
I_l(k) \equiv
\int^k_{k\eta_E/\eta _0} dy \frac{J_{l+1/2} (k-y)}{(k-y)^{5/2}}\frac{
J_{5/2} (y)}{y^{3/2}},
\ee
 $\eta_E$ and $\eta_0$ denote conformal time at recombination and at
the present era and $C(n) = P(k) k^{1-n}/4\pi$, where $P(k)$ is the
power spectrum \c{S1985}.

A very useful relationship follows after numerical integration of this
expression \c{LMM1992}. It can be shown that the ratio
\be
\frac{\Sigma_l^2[S]}{\Sigma_l^2[T]} = \frac{2}{m^2}\frac{A_S^2}{A_G^2}
= \frac{25}{2} \epsilon
\ee
is independent of $l$ if $n$ is
approximately constant and close to unity.  For potentials driven by
an exponential potential comparison with Eq. (\ref{C=0}) therefore
implies that
\be
\l{remarkable}
\frac{\Sigma_l^2[S]}{\Sigma_l^2[T]} =
\frac{\Sigma_l^2[S]}{\Sigma_l^2-\Sigma_l^2[S]} =\frac{25}{2} \left(
\frac{n-1}{n-3} \right),
\ee
where we assume that the expectations add
in quadrature, i.e. $\Sigma_l^2 = \Sigma_l^2[S] + \Sigma_l^2[T]$
[32,33]. This is a remarkable result and suggests that inflation
driven by exponential potentials can be observationally tested if the
spectral index and ratio of the scalars to tensors on large angular
scales can be determined.  Moreover, deviations from this relationship
will tell us how far away we are from the exponential potential model.

Since the large-scale Sachs-Wolfe effect is potentially due to both
scalar and tensor modes, we must consider smaller scales ($\theta
<2^o$), where the gravitational waves are unimportant, to determine
separate normalizations for these two components. However, because
these scales were sub-horizon at the surface of last scattering,
complex physical processes do not allow simple expressions for the
anisotropies to be written down. On the other hand a `Doppler peak' is
expected to occur at $l\equiv l_{DP} \approx 120-200$ due to effects
such as Thomsom scattering off moving electrons \c{BG1987}. In
principle, therefore, a measurement of the height of this peak gives
the scalars at $l_{DS}$ and comparison with the low multipole
anisotropies may allow the tensors to be separated out. If this can be
done, it would imply that the COBE satellite is indeed the first
experiment to make a direct measurement of the primordial spectrum of
gravity waves!

The dark matter in the universe affects the small-scale CMBR
anisotropy via a transfer function, although it seems that this is not
so important for degree-scale experiments.  However, the possible
effects of reionization must also be considered. To make progress it
seems that one must consider a hypersurface in the full parameter
space of observational cosmology and reconstruct within the context of
this plane. We shall therefore conclude this section by summarizing
what we feel to be the most important parameters that need to be
determined, along with some `favorite' values.

{\em 1.} $\Omega_0$: Inflation predicts that $\Omega_0=1$ and
observational support for this prediction comes from the QDOT redshift
survey of the Infrared Astronomy Satellite (IRAS) \c{iras}. Their
results suggest that $\Omega_0^{0.6}b^{-1}_{\rm I} =0.86\pm 0.15$,
where $b_I$ is the IRAS bias parameter.

{\em 2. Age of the universe}: If there is no cosmological constant,
the age of the $\Omega_0=1$ universe is $t_0=6.52 h^{-1}$Gyr. Redshift
surveys suggest that the current value of the expansion rate is
$0.4\le h\le 1.0$, but if $h>0.6$, the age of the universe is smaller
than the oldest globular clusters in the galaxy, $t_{\rm
star}=(13-15)\pm 3$Gyr. Therefore low values of $h$ are favored. The
Sunyaev-Zel'dovich effect distorts the CMBR when the electromagnetic
waves interact with the hot gas in galaxy clusters and a value of the
Hubble constant can then be determined by constraining the size of the
cluster along the line of sight. Birkinshaw et al. \c{B1991} find
$h=(0.4-0.5)\pm 0.12$ whilst Jones et al. \c{J1993} find $0.2<h<0.75$.
These observations favour the lower region.

{\em 3. Baryonic dark matter}: Limits on the percentage of baryonic
dark matter in the universe deduced from observing the primordial
abundances of the light elements are \c{OSSW1990}
\be
\Omega_Bh^2=0.0125\pm 0.0025.
\ee
Saturating the lower bound on $h$
yields the upper limit $\Omega_B\le 0.09$ and one must therefore
choose some form of non-baryonic dark matter if $\Omega_0=1$.

{\em 4. Cosmological Constant}: A number of observations suggest that
the cosmological constant may be a candidate for this dark matter
[39,40]. We note however that gravitational lensing effects imply that
$\Omega_{\Lambda} \le 0.6$ and a best fit to clustering results
implies $\Omega_{\Lambda} \le 1 -(0.2\pm 0.1) h^{-1}$, at least for
$n<1$ \c{Metal}.

{\em 5. Reionization history}: In the standard picture recombination
occurs at a redshift $z\approx 1300$ and the ionization fraction $X_e$
is given by $1+z_{\em LSS} \approx (0.03X_e\Omega_Bh)^{-2/3}$ if
$\Omega_0=1$. Thus the surface of last scattering could have formed as
late as $z_{\rm LSS}\approx 76$ if $X_e=1$, $h=0.4$ and we employ the
upper limit for $\Omega_B$. Such effects alter the position of the
Doppler peak and this implies that some assumptions about the
ionization history of the universe must be made.

It seems reasonable to assume that no reionization occurred and to
then specify $\Omega_0=1$, $\Omega_B=0.1$, $\Omega_{\Lambda}=0$ and
$h=0.5$ as a first approximation. The free parameters are then the
spectral indices of the scalar and tensor fluctuations, the relative
amplitude of the fluctuations on the quadrupole scale and the form of
the dark matter (or equivalently the transfer function).

The current state of play with the observations suggests that the most
realistic way to reconstruct is within the context of a specific dark
matter model and a constant spectral index. There are a number of
active and proposed dark matter searches in operation and one might
hope to gain some insight into the dark matter in the near future. The
most promising method of determining the primordial power spectrum is
through measurements of the peculiar velocity field. The development
of the POTENT method, in particular, is potentially very useful
because it assumes that the velocity is simply given by the divergence
of a scalar \c{BERT}.  All matter interacts via gravity, so peculiar
velocities measure the mass spectrum and {\em not} the galaxy
spectrum. In the linear regime
\be
{\cal P}_v (k) = \frac{1}{25\pi} \,
\left( \frac{aH}{k} \right)^2 \, \frac{2}{m^2} \, A_S^2(k) T^2(k) ,
\ee
where the spectrum of the modulus of the velocity $v$ is defined
as ${\cal P}_v = V (k^3/2\pi^2) \langle \left|\delta_v \right|^2
\rangle$, the Fourier components $\delta v(k)$ are defined over the
physical volume $V$ and $T(k)$ represents the transfer function. This
equation shows how the spectrum is related {\em directly} to the
amplitude of the scalar fluctuations when they reenter the Hubble
radius. In principle, therefore, the scalar spectral index can be
reconstructed.

The idea then is that the height of the Doppler peak at $l\approx
l_{DP}$ can be determined once the spectral index and transfer
function of the dark matter are known and this leads to a predicted
value for what should be observed. Eq. (\ref{remarkable}) implies that
\be
\frac{\Sigma_2^2[S]}{\Sigma_2^2[T]}
=\frac{\Sigma_{l_{DS}}^2[S]}{\Sigma_{l_{DS}}^2[T]}
\ee
and this
expression is independent of the spectral index. In principle this
gives us a measurement of the amplitude of the gravitational waves at
the scale $l_{DP}$ and should be sufficient to lift the degeneracy
intrinsic to the reconstruction procedure. Once the degeneracy has
been lifted this will allow the integration constant to be determined.

\section{Conclusion}

\setcounter{equation}{0}

In this talk we have summarized the general observational predictions
of the simplest class of basic inflationary models. The formalism that
allows one to reconstruct the functional form of the inflationary
potential from the scalar and tensor perturbation spectra was then
discussed. Full reconstruction of the potential does not appear viable
within the foreseeable future.  However, it is quite likely that an
accurate determination of the spectral index will be made within the
next few years. We presented the full class of general solutions that
lead to scalar fluctuations with a constant spectral index and it was
found that the precise form of the potential is rather sensitive to
the gravitational wave contribution to large angle CMBR anisotropies.
If $n>1$, the potential is convex and, if its functional form over the
large-scale structure range is extrapolated to the origin, it has a
global minimum located at $V\ne 0$. If $n<1$ and the gravitational
wave contribution is negligible, the field is located within the
vicinity of a local maximum instead. On the other hand, if $n<1$ and
the tensor spectrum is important, the potential takes an exponential
form.

These general solutions can be expanded as a Taylor series about some
value of $\phi$, which typically is taken to be the value of the
inflaton field when the scale corresponding to the quadrupole first
crossed the Hubble radius during inflation. This is consistent because
all large-scale structure observations correspond to only a few
$e$-foldings of inflationary expansion and, since the scalar field is
moving slowly during inflation, the relative change in the value of
$\phi$ in this range is expected to be very small. This implies that
any potential that has a Taylor expansion equivalent to these general
solutions to lowest order in $\phi$ will also produce a scalar
spectrum with constant $n$.

We therefore conclude that partial reconstruction of the inflationary
potential, within the approximation that the spectral index is
constant, is a realistic possibility. We have shown that a
determination of the spectral index, together with knowledge of the
gravitational wave amplitude on one scale, will be sufficient to
determine whether the potential is convex or concave over the narrow
range of scales associated with large-scale structure. Although this
is a rather limited piece of information, it would nevertheless
correspond to an observation of physics at the GUT scale.

\vspace{.5cm}

{\bf Acknowledgment} It is a pleasure to thank my collaborators B. J.
Carr, E. J. Copeland, E. W. Kolb, A. R. Liddle \& R. K. Tavakol who
were involved with the work discussed here.  V. B. Braginsky, P.
Coles, G. F. R. Ellis, D. H. Lyth, P. J. Steinhardt, A. N. Taylor \&
M. S. Turner are acknowledged for useful comments and discussions
regarding this work. I also express my gratitude to the organizers of
the seminar for a very enjoyable and informative conference and, in
particular, to V. M. Mostepanenko for his warm hospitality. This work
was supported by the Science and Engineering Research Council (SERC)
UK through a postgraduate research studentship and a postdoctoral
research fellowship. The author was supported in part at Fermilab by
the DOE and NASA under Grant No. NAGW-2381.


\end{document}